# ANOXIA - TREATMENT BY OXYGEN DEPRIVATION : OPTIMIZING TREATMENT TIME OF MUSEUM OBJECTS

*Michèle Gunn[1]\*, Houri Ziaeepour[2], Fabrice Merizzi[3], Christiane Naffah[4]*

*[1] Musee Quai Branly*, 55 quai Branly, 75007 *Paris, France.*

*[2] Mullard Space Science Laboratory, University College London, Holmbury, St. Mary, Dorking, RH5 6NT, Surrey, UK.*

*[3]Department of Islamic art, Louvre Museum, 34 quai Louvre 75001 Paris, France*

*[4] Centre de Recherche et de Restauration des Musées de France 14 Quai François Mitterrand 75001 Paris, France.*

## Abstract

ANOXIA, treatment by oxygen deprivation is largely used for decontamination and disinfestation of cellulose and protein-based organic materials. More specifically this method is applied to more than one hundred thousand of objects destinated for a new museum in Paris, "Musee du Quai Branly". We describe the anoxia installation in this museum and report the result of a study regarding the efficiency of this method and the optimum treatment time, crucial for treating a large collection. We show that the standard 21 days of exposure is not always the optimal choice. Temperature plays a crucial role for hastening the death of insects found within objects. At a temperature of 25°C, it is entirely possible to reduce exposure times to 10 or 15 days for the insect species commonly found in museums. The oxygen drop times is between 1 and 2 days for most objects, depending on type and porosity of materials. This corresponds to a treatment time between 15 and 16 days. The effect of humidity is less clear. It can increase the necessary treatment time both for larvae and for adult insects.

---

\*    Correspondence should be addressed to M.G. (mgu@quaibranly.fr)

*INTRODUCTION*

A large part of the collection of the future Musée du quai Branly - currently under construction near the Eiffel Tower in Paris - is composed of cellulose and protein-based organic materials. Such materials are favourable media for the development of micro-organisms and insects, leading to their degradation.

This collection is presently being treated in a series of steps which include cleaning, the taking of photographs, packaging and biological decontamination in the Le Berlier building which has been especially equipped for this purpose.

The collection of the Musée du quai Branly, numbering about 275000 objects[1], comprises on the one hand collections from the Musée National des Arts d'Afrique et d'Océanie (MNAAO) and the Musée de l'Homme (MH), and on the other hand has been enriched by new acquisitions.

Studies of the general state of conservation of these collections in their original institutions by experts, demonstrated the existence of infestation by *Anobiidae*, *Dermestidae* and *Tineidae*, to name just a few. Infestation was found to be more or less serious depending on the institution and departments in question. Given that it is difficult to reconstruct an accurate case history of the infestation and the steps that have been taken to counter it, it was decided to proceed with treatment of all objects containing organic materials, without exception.

This prudent choice was made in view of the fact that the treated objects were not destined to return to the site from which they came, but were going to be housed in a new museum: a « complete overhaul » of the objets in order to reduce the level infestation to zero is advisable under such circumstances. Furthermore, the objects are treated by oxygen deprivation (anoxia), which minimises the risk of chemical degradation, although some discolorations of some pigments have been reported (TOSHICO, K., 1980); this cannot be said of classical fumigation treatments even though the treatment times are much shorter in the latter cases.

Heritage institutions currently employ oxygen deprivation treatment times ($T_t$) of 21 days. This duration appears to have been adopted in the light of the results of experiments carried out on a particularly resistant species, the rice weevil, an important pest in the **food industry** : 500 hours (**21 days**) at **26°C, 12% relative humidity, in a nitrogen atmosphere containing 1% oxygen**. The exposure time is extended to 1000 hours (6 weeks) if the temperature is lowered to 20°C (SELWITZ, C. *et al*, 1998).

Of more relevance in the **museum field**, the old house borer, *Hylotrupes bajulus*, is also a species resistant to treatment by oxygen deprivation. Its favourite medium is resinous wood. Eradication of this insect necessitated 20 days in somewhat different conditions: 20°C and 40% relative humidity. The duration can be reduced to 10 days if the temperature is raised to 30°C (VALENTIN, N., 1993).

It has gradually become standard practice to use a treatment time of 21 days. The recommended conditions are in general as follows: less than 0.1% oxygen, a temperature above 20°C and relative humidity of 50%.

---

[1]    The exact number will be established at the end of the collection treatment programme

There are a very large number of objects to be treated (more than 80% of the collection). The deadline for completion of the collection treatment programme leads to constraints, in view of which time is of the essence. It is therefore appropriate to analyse the time given to each stage of the object treatment process in order that none be wasted.

If a reduction in the duration of oxygen deprivation treatment turns out to be possible, this would enable a good speed to be maintained during the progress of the collection treatment programme.

Consequently the key conclusion awaited from this study is the answer to the following question: is the anoxia treatment efficient for an exposure time less than 21 days ?

Each anoxic treatment installation has its own characteristics. Thus, since the installation we have used, named EPMQB (named according the name of the musée, Etablissement Public Musée du quai Branly[2]), was specially designed for the Musée du quai Branly treatment site, and was of a new and as yet untried form in the field of heritage and conservation, it was necessary to carry out a study in order to optimize the conditions of treatment for objects, in particular as regards oxygen drop times and exposure times.

The feasibility of treating infested museum objects by oxygen deprivation, either through the use of oxygen scavengers, or in a controlled atmosphere of an inert gas such as nitrogen ($N_2$) or argon (Ar)) or carbon dioxide ($CO_2$) is now well established. Resistant species such as *H. bajulus or A. punctatum* (cellulose) can be totally eradicated, and this is also possible in the case where insects are at the egg or larval stage which renders them more resistant to treatment (Rust, M. et al, 1996 ; Selwitz, C. ; Maekawa, S., 1998). Many studies have already been carried out by teams in the USA (Getty Conservation Institute) and Australia (Australian Museum) for exemple. These studies have enabled the evaluation of influence of different parameters, such as the level of oxygen ($O_2$), temperature and relative humidity, on the exposure times needed to achieve 100% mortality whatever the life cycle stage of the insects.

Therefore, the goal of the study is to determine the efficiency of the EPMQB equipment and the effectiveness of the traitement in the case of insects buried deep within an object. This phase of the study should enable the degree to which oxygen is removed from the inside of treated objects to be evaluated.

We report and discuss results obtained in the following areas :

1) the exposure time, $T_e$, in the new EPMQB installation, leading to 100% mortality irrespective of life cycle stage of the insects present in infested objects. The treatment conditions are based on previous literature reports. They must be optimized from a mortality viewpoint whilst avoiding endangering at the same time the physical structure of the treated objects: an atmosphere with highly reduced oxygen content is used, between 1000vpm and 30vpm, a temperature of 25°C±1°C and hygrometry of 50%±5%.

2) the oxygen drop time, $T_i$[3], defined as being the time taken to lower the oxygen content in the treatment unit to the required level (0.1 %), and to study the effect of the degree of loading with museum objects.

3) the oxygen desorption time of the objects $T_d$. The $T_d$ value depends intrinsically on the nature of the materials and the volume of the objects to be treated and the volume of the

---

anoxia chamber. The $T_d$ varies as a function of the permeability of the materials to gases, i.e. nitrogen and oxygen in this case.

---

**A - PROGRESS IN OXYGEN DEPRIVATION TREATMENT: MAIN RESULTS OBTAINED BY OTHER INSTITUTIONS**

---

## I-TESTS ON INSECTS: EXPOSURE TIMES AND CLIMATIC CONDITIONS OF TREATMENT

The experiments carried out cover a very wide range of insects at all life cycle stages, i.e. eggs, larvae, nymphs and adult insects. Atmospheres were modified through the use of the three most frequently used gases: carbon dioxide ($CO_2$), nitrogen ($N_2$) and argon (Ar). The anoxia treatment was performed in bubble chambers (VALENTIN, N.,1994; RUST, K. *et al.*, 1996). In some cases in order to simulate their being buried, insects were prepared in glass tubes closed by a system allowing gaseous exchange. The tubes were subsequently left fixed within blocks of wood. Other experiments were with sections of pine wood and with books of dimensions which were artificially infested with the insects to be studied. The main results show that :

- the most resistant life cycle stages of insects are eggs and larvae;
- not all the insects react in the same way, the old house borer is the most resistant;
- the treatment is more effective with argon than with nitrogen;
- temperature is an important factor whatever the other conditions.

The exposure time is reduced when the temperature increases; for example in the case of *H. bajulus* (old house borer), when the temperature increases from 20°C to 40°C, the exposure time is reduced from 21 days to 2 days. The studies carried out in **the museum environment** show exposure times much lower than the standard of 21 days when the climatic conditions are chosen appropriately, including for the most resistant species; for example :

- for the *Anobiidae* (e.g. : furniture beetle), complete elimination was achieved after **3 or 5 days** with 50% relative humidity, 30°C et 0.03% oxygen**.** One exception to this range has been reported; this is Lasioderma serricorne (cigarette beetle), which required **8 days** at 25°C with 50 % relative humidity, or **9 days** at 20°C and 40 % relative humidity. (VALENTIN, N., 1993);

- for *Hylotrupes bajulus* (old house borer) of the *Cerambycidae* family, which has shown itself to be rather more resistant, the exposure time was **10 days** at 30°C with 40 % relative humidity or **20 days** at 20 °C. (VALENTIN, N., 1993);

- for the *Tineidae* (e.g. clothes moths), 4 days at 25.5°C, 55% relative humidity and an oxygen level below 0.1 % have been reported (RUST *et al*, 1996).

More recently in 2000, a Japanese team proposed a practical protocol for anoxia treatment according to the type of insect. They advise 25°C or 30°C for one to three weeks. When the temperature is 20°C exposure time should be extended to 10 weeks with an oxygen level of 0.2%. In evidence, in this case the high level of the oxygen makes the treatment time longer than in the previous experiments (KIGAWA, R. *et al*, 2000).

**The experimental parameters for treatment, according to the infestation, are actually well known. The key questions which now remain to be answered are whether or not the experimental conditions are really achieved within the treated objects.**

## II-DEMONSTRATION OF OXYGEN DESORPTION OF TREATED OBJECTS

The results outlined above were obtained by simulating the burial of insects within objects in order to mimic as closely as possible a real situation. However, it is difficult from a technical point of view to directly evaluate the desorption time $T_d$ of oxygen from objects, since to do this would in principle necessitate the positioning of detectors within the objects. $T_d$ depends on the porosity of the objects studied, and on the permeability of the materials to the gas used. The duration of treatment therefore depends on the desorption time. Various methods were used to evaluate desorption such as calculation of the time needed for the oxygen level to reach its equilibrium value, and comparison of the treatment times of infested objects with the treatment times of reference samples.

**II-1 calculation of the  time needed for equilibrium to be reached.**

Simulation was carried out with **fresh, non-infested wood** from different sources: poplar, oak, walnut, having fixed dimensions. The wood samples were bare or covered by a thick protective coat. They were enclosed in a pocket of volume 32 litres ($0.032m^3$) until equilibrium was reached. The initial experimental conditions were:
23°C, 0.1% to 0.2% oxygen. Equilibrium was reached with 0.4% oxygen. Oxygen desorption was found to be more difficult with painted wood than bare wood. The difficulty increase in the way poplar, oak, walnut, with a maximum desorption time of 120 hours (5 days).
The same experimentrepeated for **infested walnut wood** showed a shorter desorption time: equilibrium was reached more quickly, in one hour for the bare wood and in **4 hours for the painted wood**. This result is a consequence of the greater porosity of infested wood in comparison to healthy wood on account of the tunnels hollowed out by the infesting insects (SELWITZ, C., 1998).

**II-2 Comparison of treatment time of infested objects with treatment time of reference samples (test samples)**

Treatments carried out on museum objects using readily available test samples as references and performed in a controlled argon atmosphere showed similar exposure times for the objects and the reference samples, the differences in exposure times being one day or less.

This result was not always observed. Wooden objects (pianos, sculpture and panels of wood) infested by *Anobium punctatum* (furniture beetle) had to be treated over 10 to 14 days, compared to only 4 for the reference samples. Similarly, 7 days were needed for a textile sample of dimensions 135x87x43cm infested by *Attagenus megatoma* (black carpet beetle) as compared to 2 days for the corresponding reference sample. The treatment conditions were as follows : a temperature above 20°C, relative humidity of 40 % to 50 % and an oxygen level of **0.02-0.04%** (VALENTINN, N., 1993).

Other experiments performed on powderpost beetle and termites buried in the heart of the wood and sealed and in other cases non-sealed (more accessible) did not show any difference in exposure times (RUST, KENNEDY, 1993).

## III- DETERMINATION OF TREATMENT TIMES FOR WOODEN OBJECTS

The results below (table 1) were obtained for large objects infested by *A. punctatum* (furniture beetle) and *H. bajulus* (old house borer).

Table 1 : Time and treatment conditions for museum objects using **Argon.**
According to Nieves Valentin, 1993

| Objects | Dimensions cm | Insects present | Temperature °C | Relative humidity % | Oxygen level (%) | Exposure time (days) |
|---------|---------------|-----------------|----------------|---------------------|------------------|----------------------|
| Piano | 200x100x100 | *A. punctatum* | 25 | 40 | 0.03 | 14 |
| Panel | 175x64x35 | *H. bajulus* | 20 | 50 | 0.04 | 15 |
| Sculpture | 200x80x52 | *A. punctatum* | 25 | 45 | 0.04 | 10 |
| Frame | 75x45x15 | *H. bajulus* | 20 | 45 | 0.03 | 10 |

This table shows exposure times $T_e$ between **10 days and 15 days.**

## B -THE STUDY CARRIED OUT AT THE MUSEE DU QUAI BRANLY

The bibliography cited above shows that the most resistant insect developmental stages are eggs and larvae; adults and nymphs are the first to be affected by oxygen deprivation. *Hylotrupes bajulus* (common name: old house borer), has turned out to be the species most resistant to oxygen deprivation treatment. This insect is however not very commonly encountered in museum objects. It is generally found instead in resinous wood structures of buildings. It was nonetheless chosen for the present study on account of its resistance. This insect was also readily available, being bred prior to this study at the Centre Technique du Bois et de l'Ameublement, CTBA (Technical Center for Wood and Furnishings).

At the same time as carrying out experiments on the insects, other experiments were performed in order to evaluate oxygen desorption times from the materials used, the degree to which the oxygen content within the materials was lowered and the effect of the type of objects loaded on the oxygen drop time.

# I-METHODOLOGY

## I-1 The installation used[4] : EPMQB ANOXIA SYSTEM

### I-1-1 Description of the installation

**The anoxia system equipment has five parts**

**1)** *A Pressure Swing Adsorption (PSA) nitrogen production unit (TechnicAir) composed of **a** an air compressor, **b** an air dryer and submicronic filter system, **c** a compressed air reservoir (1m³), **d** a nitrogen/oxygen separation subunit made of two receptacles containing activated charcoal molecular sieves (CMS), **e** a vessel containing distilled water for humidifying the nitrogen, **f** an oil-water separator to avoid discharging insoluble compounds into the general waste water sewage system.*

**2)** *A nitrogen storage unit with four nitrogen reservoirs each having a capacity of 3m³*

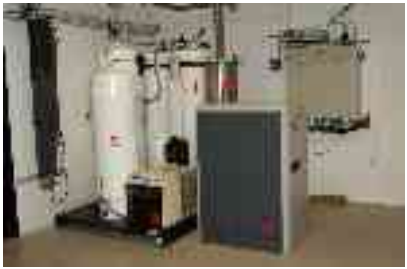 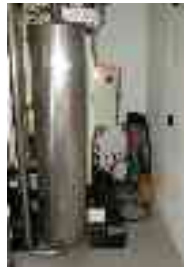 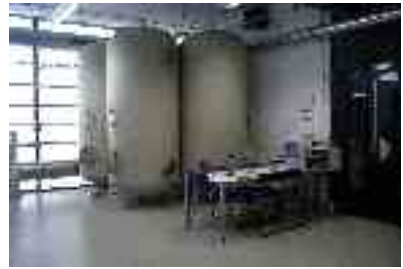

Nitrogen production unit            Reservoirs of nitrogen

**3)** *A treatment unit with three rigid containers A, B, C having a volume of 1x25 m³ (B) and 2x35 m³ (A, C).*

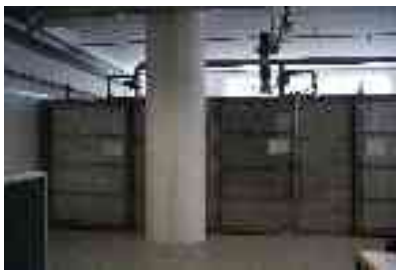 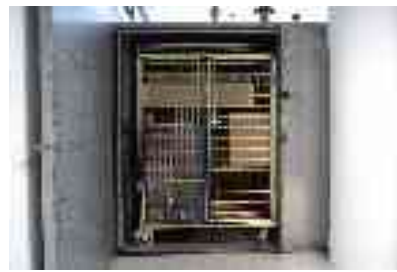

Three rigid containers A, B, C            View of the loading in the container

4) *A remote control system enabling control and monitoring of treatment containing: an electric command desk housing a TSX 37 system sold under the name of Télémécanique-Schneider, a PC computer connected to a printer, which enables the reading and recording of the treatment parameters throughout the treatment cycle.*

---

[4]    This installation was built by the company Mallet (division of CATS)

## I-1-2  System operation

The technique chosen for oxygen deprivation is a dynamic system based on a continuous flow of nitrogen through the enclosed treatment units. The nitrogen used is prepared from the air in the room. The nitrogen is separated from the oxygen by the molecular sieve system. The nitrogen is stored in a series of reservoirs. Oxygen is desorbed from the molecular sieves under the pressure of nitrogen.

Treatment protocols are entered by keyboard and recorded in a computer file reserved for these data: degree of humidity: 50%, exposure time:14 days, level of oxygen: 1000 vpm (0.1%). These values can be changed. The temperature of the enclosed treatment units is the same as that of the surrounding area, e.g. 25°C.The humidification of the enclosed treatment units is performed through the humidification of the nitrogen.

## There are three steps to the treatment cycle:

1) a purging phase of the enclosed units called *"INJECTION GAZ 1"* and *"INJECTION GAZ 2"*, which reduce the oxygen level  to below 0.1%;

2) a treatment phase with an oxygen level below 0.1%, called "CONTACT GAZ";

3) a "rinsing" out phase : the oxygen level is raised to 20% by sucking in air from the room, called "RINÇAGE", followed by the end of the treatment cycle, called *"FIN TRAITEMENT"*.

## The oxygen drop times depend on the volume of the enclosed treatment units and on the nature of the objects loaded within them.

*As a matter of definition, the exposure time $T_e$, is considered to start once the oxygen level has decreased to 0.1% at beginning of the phase called "CONTACT GAZ".*

*All the parameters are recorded and stored in files which can be accessed using Excel.*

*Averages are calculated on the basis of these data. At the end of a treatment cycle, a report is produced. The latter displays the dates and times of the following events: beginning of the treatment, moment when oxygen level reaches 0.1% (called "CONTACT GAZ"), end of the treatment.*

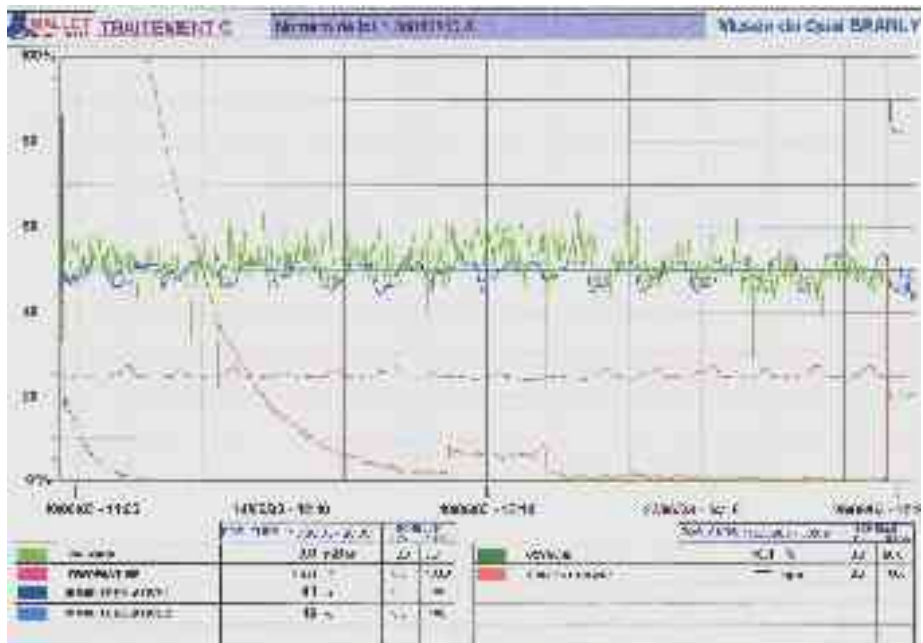

**Curves of parameters throughout the cycle of treatment**

## I-2 Looking for the proper exposure time: experiments on insects

Four series of experiments were performed with *Hylotrupes bajulus* at egg and larval stages with the aim of determining the lowest exposure time. Each series was composed of three experiments with identical treatment times. The eggs are placed on blotting paper and conditioned in Petri dishes. Larvae are placed in small wooden blocks and enclosed in Petri dishes.

Each experiment was carried out in the following manner : three egg test samples and three larvae test samples were placed in the oxygen deprivation sealed unit. One reference egg test sample and one reference larvae test sample were left in the atmosphere of the area outside the enclosed treatment units. The first series were treated using an exposure time $T_e$ of 10 days. This is the shortest time given in the existing literature. After the experiments, the test samples were put back in steamroom for 15 days. The % mortality arising from the oxygen deprivation treatment can then be calculated after counting dead eggs and larvae and also survivors. Results are shown in table 2

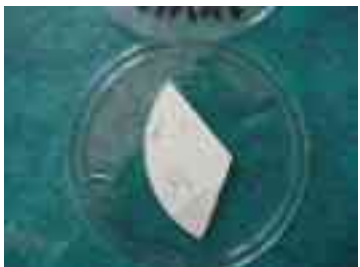

**Egg test sample**

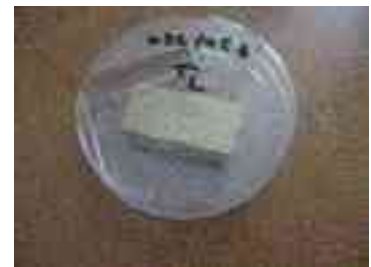

**larvae test sample**

**Table 2. Mortality rates for exposure times of 10, 7 and 5 days.**
**H. bajulus (old house borer)**

| Experiment | Exposure time (days) | Test samples in nitrogen atmosphere Mortality rate % | |
|---|---|---|---|
| | | Eggs | Larvae |
| Series 1 | 10 | 100<br><br>no hatching | 100 |
| Series 2 | 7 | 100<br><br>no hatching | 100 |
| Series 3 | 5 | 64<br><br>hatching | 100 |
| Series 4<br><br>Samples in plastic boxes | 14 | no hatching | 100 |

In the case of series 4, samples conditioned in Petri boxes were enclosed in sealed plastic boxes to mimic the deep burying of insects.

**The lowest exposure time at 25°C, 50% RH, with an oxygen level below 0.1% is 7 days**

**I-3 Oxygen drop times $T_i$ and desorption time $T_d$ in the treatment unit**
These values were calculated in the following manner, based on the automatic treatment report :

$T_i$ = (time at which oxygen level of 0.1% is reached, called "CONTACT GAZ") - (time of beginning of treatment).

Desorption times $T_d$ are determined by comparing the $T_i$ observed when the containers are loaded with objects to be treated with the $T_i$ observed when the containers are empty

**$T_d$ = $T_i$ (loaded unit) - $T_i$ (empty unit)**

*Table 3: Average desorption time for unit A, B and C*

| Unit | Average $T_i$ of treatment units when loaded | Average $T_i$ of empty treatment unit | $T_d$ |
|---|---|---|---|
| A 35m³ | 1day + 9 hrs ±2hrs | 23 hrs | 10 hrs |
| C 35 m³ | 1day + 13 hrs ±2 hrs | 22 hrs | 13 hrs ±2 hrs |
| B 25m³ | 1day ±1hr | 13 hrs | 11 hrs ±1hr |

The $T_i$ and $T_d$ averages were calculated on the basis of six loadings. The load is analysed, batch by batch and object by object, using TMS (The Museum System) files, where information concerning the materials of which the objects are made, as well as their dimensions and weight, are recorded.

It appears possible to infer from the composition of the different loads that a load made up principally of wooden objects or wooden object and textile or skins, requires a priori a longer oxygen drop time than a load of wood and vegetal pulp or vegetal fibres. However differences in the space occupied by objects appeared to be the dominant factor. When the volume of the closed treatment units is most effectively filled, the oxygen drop time is longer.

In contrast, it became clear that the sealed treatment units A and C, which each have a volume of 35 m³, reach the stage called "CONTACT GAZ" (oxygen level reduced to 900vpm) after a day and a half whereas the treatment unit B, having a volume of 25 m³, only required one day.

## II-Demonstration of the diffusion of nitrogen

### II-1 Through the wrapping

Diffusion of nitrogen has been demonstrated by comparison of the oxygen level in a cardboard box with the oxygen level in the sealed treatment unit.

Table 4 : Time taken for homogenization of atmospheres between cardboard packing boxes and the sealed treatment units (here unit A)

| Place of oximeter | Initial oxygen level (12h) | Oxygen level after 15 minutes | Oxygen level after 40 minutes | Oxygen level after 5 hours (16h55) |
|---|---|---|---|---|
| Treatment unit A | 21 | 13.6 | 2.6 | 4.5 |
| Cardboard box | 21 | 20.6 | 12.8 | 4.5 |

The oxygen level is the same in the cardboard box and in the treatment unit about five hours after the beginning of the treatment cycle.

*The same phenomenon of equilibrium, here concerning oxygen concentration, occurs between the treatment unit and the treated objects. Splits, cracks, tunnels dug out by insects, and natural micropores in the wood can be seen as analogous to the opeining cut out in the cardboard box.*

In order to study how oxygen concentrations vary within treated objects, a simulation was carried out. Two wooden cases were used[5], one in pine and the other in laminated wood, which is less porous. Both had an interior volume of about 3 litres. Such a volume is a prerequisite for the removal of air samples for later analysis. The pine case had a thickness of 7 cm all around the empty internal space. The laminated wood case had a thickness of 1 cm.

### II-2 Demonstration of the degree of oxygen reduction within objects : experiments on gas diffusion simulation with wooden cases

The degree of oxygen reduction within objects was studied through experiments carried out in treatment unit A which is equipped with two probes for measuring oxygen levels.

- **Test 1**: Carried out on a pine case of internal volume 3.1 litres and of thickness 7 cm. The valve (V2) linking the case to the oxygen-analyser was disconnected at the beginning of the experiment so as not to interfere with gaseous exchange (of nitrogen and oxygen). The other valve (V1), which enables oxygen levels to be measured in the treatment unit, remained connected to the oxygen-analyser.

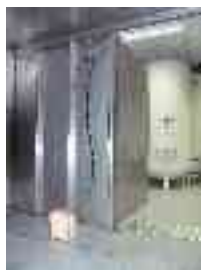 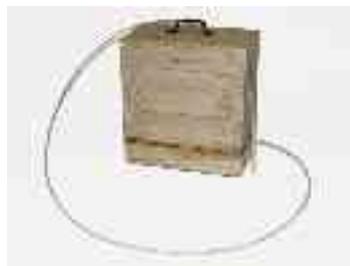

**Three days after** the beginning of the treatment cycle, V2 was connected. The oxygen level inside the box was measured after having calibrated the oxygen-analyser with the nitrogen produced. The oxygen level was **0.0337%** (337 vpm).

**Five days after,** the oxygen level was **0.0010 %** (10 vpm).

- **Test 2**: carried out on a pine case during a test of unit A (a) when not loaded with objects, (b) when loaded with objects.

Oxygen levels in the box were measured every 24 hours for three days. The results are shown in table 2.

---

[5]    The wooden cases were made by the  company Hygiène Office

*Table 5 : Change of oxygen level in the pine box compared to oxygen level in treatment unit*

| Location of the oxygen trace analyser | (a) Oxygen levels (vpm) Unit treatment without load | | | | (b) Oxygen levels (vpm) Unit treatment with load and pine box sealed with a plastic film | | | |
| --- | --- | --- | --- | --- | --- | --- | --- | --- |
| | 24 hours | 48 hours | 72 hours | 86 hours | 24 hours | 48 hours | 72 hours | 86 hours |
| Unit A | 786 | 496 | 354 | 358 | 2743 | 1090 | 708 | 370 |
| Pine box | 891 | 574 | 425 | 348 | 9207 | 2709 [6] | 1191 | 577 |

Figure 1a: the evolution of oxygen drop in the pine box sealed with plastic film starting from 21.5 %

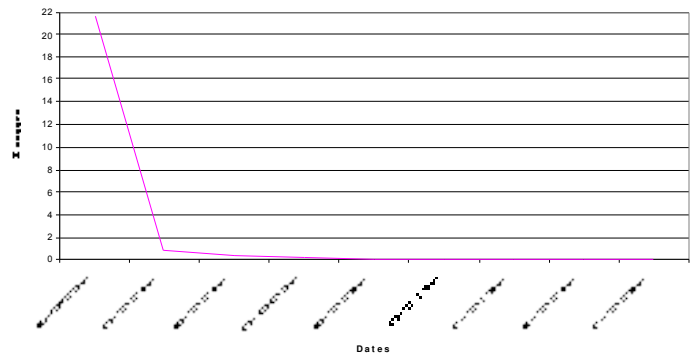

Figure 1b: blow  up of the low part of figure 1a

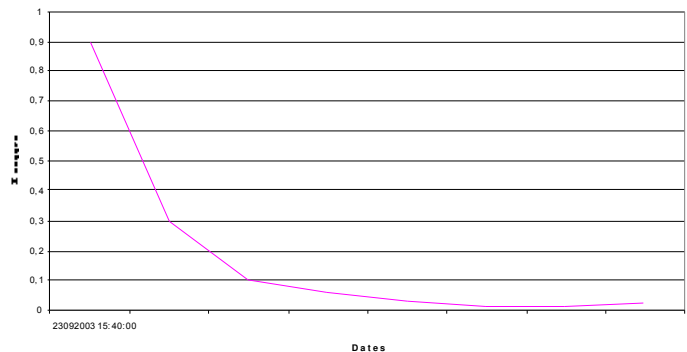

[6] In the case of a pallet of objects sealed with a plastic film this value is 7000. Figure 3 compares oxygen levels in the sealed pallet and the treatment unit. The great difference in the drop of oxygen tends to disappear as the treatment progresses.

Figure2 compares oxygen levels in the sealed pallet and the treatment unit. The great difference in the drop of oxygen tends to disappear as the treatment progresses.

Red curve : sealed pallet.   Black curve: treatment unit

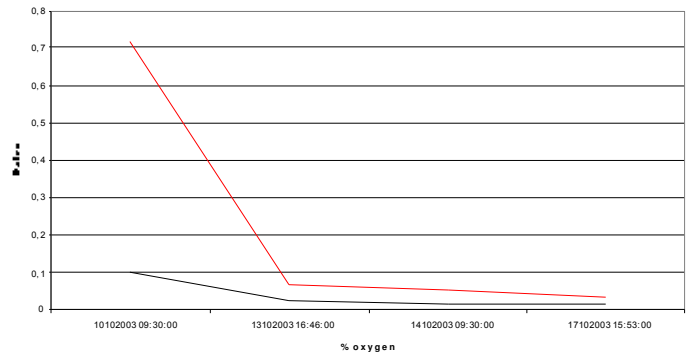

Figure 2: Comparison between the palette and the treatment unit

Figure 3 : The drop of oxygen in the treatment unit A

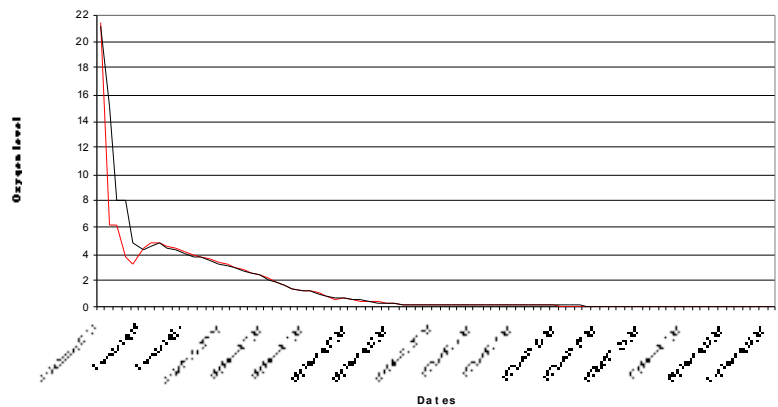

Figure 3 : The drop of oxygen in the treatment unit A

The oxygen level is measured at two point inside the treatment unit symmetric with respect to the gas extractor. The small difference is due to asymmetric extraction of gas with a rotating fan.

We interpret the slight increase of the oxygen level after around 2 hours to be due to injection of oxygen from objects inside the treatment unit: due to extraction of gas,  the partial pressure of oxygen in the treatment unit becomes smaller than the oxygen pressure inside the porous material of objects under treatment, consequently oxygen begins to be extracted from objects and for a short time a increase in oxygen level is observed. However as the volume of the oxygen inside the pores is very limited, the effect is small and will disappear quickly and the exponential decrease of oxygen take over again.
This observation is very important because it proves in this process that not only nitrogen diffuses in the porous material but also the latter loses its oxygen which helps to kill larvae buried deep inside objects.

These figures show that the drop of oxygen depends not only on the time but also on the porosity of materials and on the size of the objects. The different data observed as regards desorption times, all suggest that oxygen desorption is rapid with the EPMQB anoxia installation. Thus, eradication of *H. bajulus*, old house borer, requires a shorter exposure time than those described in the literature: 7 days as compared to 10 days, in a reduced oxygen atmosphere (less than 0.1%) with a temperature around 25°C and about 50% relative humidity.

Desorption times of 10 to 13 hours are achieved, for all the enclosed treatment units used, for a target oxygen level of 0.09%. Desorption is concurrent with the oxygen drop in the sealed treatment unit so that when the stage called *"CONTACT GAZ"* is reached, the same conditions are reached within the most materials, as is confirmed by test 2 (a). For the less porous ones, *"CONTACT GAZ"* is reached 1 or 2 days later. Homogenization between the sealed treatment unit and the cardboard packing box is achieved after 4 hours. This means that during the 9 remaining hours (13-4), evacuation of oxygen from the most inaccessible spaces in the unit is taking place, e.g. in side the objects.

**III-DISCUSSION**

The tests on insects and the tests of gas diffusion cannot be performed on real objects. Only simulations are possible. Such simulations must be as close as possible to a real treatment situation or mimic the most difficult situations likely to be encountered. To this end, we decided to leave closed Petri dishes in the treatment unit, placing them in cardboard boxes under silk paper.

For the same reasons, the boxes used as models for objects to be treated during the diffusion studies either had a thickness (7 cm) greater than generally found for objects to be treated or a thickness close to that of treated objects but less permeable (laminated wood, or wrapped with plastic film).

In these artificial but tough conditions, it appears reasonable to extrapolate the results, in so far as the results obtained are reproducible and agree with those known in the literature. These conditions appear to have been satisfied, our results being constant.

The conditions leading to the death of insects are now well known; they are inescapable and in no doubt from a scientific point of view. According to the type of equipment used, these mortal conditions are attained within treated objects after variable amounts of time. Treatment *stricto sensu* only begins when those conditions are reached.

In the case of the EPMQB installation, the desorption phenomenon is observed both by an increase in the oxygen level in the curve showing the evolution of oxygen levels over time and in an indirect way by comparing the oxygen drop times of empty treatment units with those of loaded treatment units. The longer oxygen drop times observed for loaded treatment units are not simply a consequence of the modifications in gas circulation on account of large obstacles (trolleys, boxes), but also result from the phenomenon of oxygen desorption.

We consider that in our system we have a **binary mixture**, nitrogen and oxygen. Hydrosoluble pollutants such as $CO_2$ and $SO_2$ are essentially eliminated with the water in the desiccator. Water vapour is constant and does not therefore influence gaseous $O_2/N_2$ exchange. The parameters of temperature and pressure are far removed from conditions necessary for the liquefaction of gases. It is therefore reasonable to consider the $O_2/N_2$ mixture as a **perfect gas.** The mixture thus obeys the perfect gas law **PV = nRT** (P=pressure, V=volume, n= number of moles , R= perfect gas constant, T= temperature in Kelvin). The process, although dynamic, can be split up into a series of equilibrium states established between the bulk volume of the sealed treatment units and the boxes (or objects). Following **Lechâtelier's principle**, a system in equilibrium moves from one equilibrium state to another equilibrium state in such a way as to compensate for any perturbation imposed upon it. Here, the perturbation is the disappearance of oxygen from the sealed treatment unit.

When a substance disappears, the equilibrium shifted in such a way as to favour the appearance of that substance:

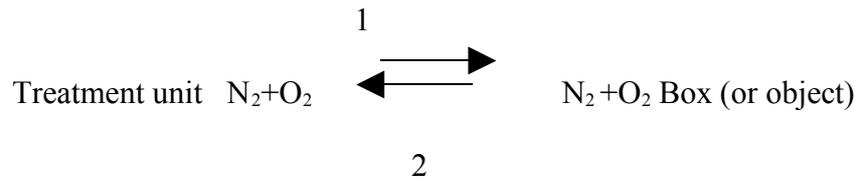

When oxygen levels drop in the treatment unit, the equilibrium is shifted in direction 1 so as to equalize oxygen concentrations.

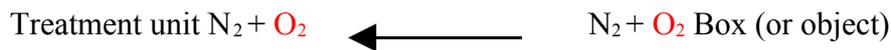

The treatment unit – cardboard box equilibrium also enables us to study the progress of the levels of nitrogen and oxygen.

Table 6 : Partial pressures of gases

| *Position of oximeter* | *Initial oxygen level* | *Oxygen level after 40min* | *Partial pressure of oxygen* | *Partial pressure of nitrogen* | *Direction of migration of O²* | *Direction of migration of N²* |
|---|---|---|---|---|---|---|
| *Sealed treatment unit A* | *20* | *3* | *0.03P* | *0.97P* | ↑ | ↓ |
| *Cardboard box* | *20* | *12.8* | *0.13P* | *0.87P⁵* | | ↓ |

*P is the total pressure in the treatment unit, and is close to atmospheric pressure. It is recorded periodically by an automatic measuring device. The partial pressures are calculated using the law of perfect gases. The partial pressure of oxygen in the cardboard box is 4.5 times greater than the partial pressure of oxygen in the treatment unit. Oxygen molecules thus naturally move in the direction cardboard box-treatment unit.*

*Modelling of nitrogen diffusion in porous objects enables the extrapolation of the findings from this study to other materials and the prediction the oxygen desorption time according to the material type (e.g. wood, feather, skin, etc.) and the size of the objects and the estimation of the effectiveness of anoxia at the core of a treated object. We consider the dispersion media to be constructed from a series of microscopic connected tubes through which the fluid passes and disperses (BEAR, J. 1963). The duration for passing through each micro-tube is random as well as the number of tubes visited by a tracer to arrive to a given depth in the media. The distribution of passage time through N tubes for large N is Gaussian:*

$$C_N(t) = \frac{C_0}{\sqrt{2\pi N}} e^{-\frac{(T-t)^2}{2N}} \qquad T = \frac{(t - \Delta t N)}{\tau} \qquad (1)$$

$C_0$ is fluid density at contact surface; $?t$ is a delay time that fluid spends in a tube before it begins its dispersion, $\tau$ is the characteristic dispersion time which depends on the microscopic interactions and viscosity of the fluid. Defining an average tube length:

$$\bar{\Delta l} = \bar{x_t}/N$$

distribution (1) changes to:

$$\frac{C_N(t,x)}{C_0} = \mathcal{N}(x) \exp\left(-\frac{\frac{t}{\tau}\left(\frac{t}{\tau} - \frac{\Delta t}{\tau}\right)}{\frac{2x}{\Delta l}}\right) \qquad (2)$$

$$\mathcal{N}(x) = \sqrt{\frac{\Delta l}{2\pi x}} \exp\left(-\frac{x^2}{\Delta l^2 \tau^2}\right) \qquad (3)$$

$$\bar{v} = \frac{\Delta l}{\tau - \Delta l} \qquad (4)$$

For $?t = 0$, (2) is simplified to:

$$\frac{C_N(t,x)}{C_0} = \mathcal{N}(x) \exp\left(-\frac{\frac{t}{\tau}\left(\frac{t}{\tau} - 1\right)}{\frac{2x}{\Delta l}}\right) \qquad (5)$$

When $t/\tau >> 0$, (5) is a Gaussian with $\sigma^2 \qquad \frac{x\tau^2}{\Delta l}$.

All the microscopic phenomena are concentrated in the quantity:

$$Q = \frac{\tau^2}{\Delta t}$$

which must be determined experimentally. For the simple geometry we have considered here i.e. a media with infinite volume and infinite contact surface with fluid, $\tau^2$ is proportional to $x$, the dispersion depth. For a finite object the dependence must be more complicated and should also depend on the dimensions. Nonetheless, the test with a box (Fig. 3) shows that at least for relatively large objects Gaussian distribution is very good fit. Dependence on $x$ however must be verified experimentally.

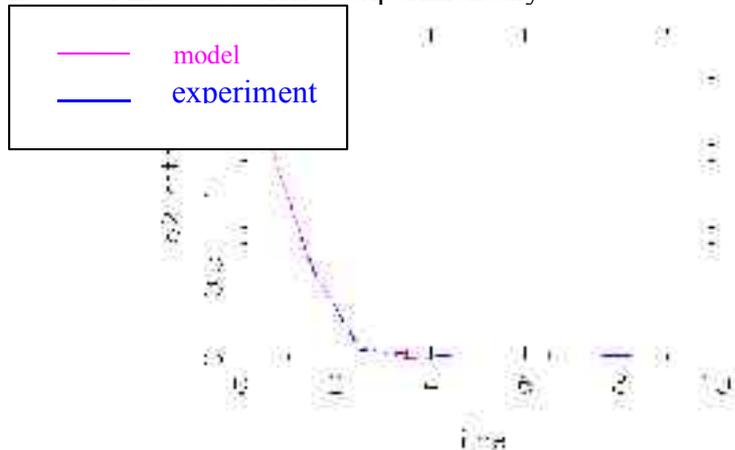

*Fig. 4: Evolution of oxygen fraction inside the box.*

This calculation will enable us to predict the oxygen desorption time within the objects according to material (e.g. wood, feather, skin) and the size of objects and establish the effectiveness of anoxia at the core of a treated object.

**Another key question is whether the treatment conditions chosen enable the eradication of any mutant insects present which may have become resistant to oxygen deprivation.**

Several workers in the field have already looked into the question of whether insects can adapt to modified atmospheres with very low oxygen content. They have shown that it is not impossible for insects to survive without oxygen but this is very rare, and encountered above all in the case of aquatic insects having water or ice habitats, for example certain small flies living in water. The latter are able to resist more than 100 days in the absence of oxygen by slowing down their respiration, increasing the amount of stored oxygen and also by a physiological change which allows them to avoid drying out. In very humid conditions, even in a reduced oxygen atmosphere, desiccation, the primary cause of death for insects, does not occur (ZEBE, E. 1991).

However, in conditions which are far removed from aquatic conditions, such as in the food industry, or in heritage institutions, such an adaptation is unlikely given that the relative humidity is much lower. Resistance may arise in the case of insects which develop in confined and very humid spaces where the oxygen concentration is low and the carbon dioxide concentration is high, or with insects buried inside books or wood. A known example is that of *L. serricorne*, the cigarette beetle. In such a case, it is sufficient to lengthen the exposure time (SELWITZ, C., 1998).

A study of the resistance of *Tribolium castaneum* (Red flour beetle) to a reduced oxygen atmosphere was carried out using the following conditions: 0.5% oxygen in nitrogen, or 20% oxygen and 15% nitrogen in carbon dioxide with 95% relative humidity, these conditions being maintained until only 30 to 50% of the insects were still alive. The surviving insects were bred for 40 generations in the same conditions. This study showed that these insects were only resistant towards the specific atmosphere to which they had been subjected. Insects returned to normal air after 13 generations displayed continuing resistance for 8 generations. The authors of this study considered that adaptation to atmospheres modified with carbon dioxide is more likely to occur than in the case of modification with other gases. More recently, in 2001, HOBACK and STANLEY studied several microhabitats where insects are under hypoxia or anoxia such as stored grain or decaying wood. In such conditions insects reduce their respiration rates when the oxygen level reached 10 %. Certain insects as *C. vomitoria* larvae can survive at 1 % only 5-6 days. However, in general, insects cannot withstand an oxygen level of 0.5 % for a long period of time.

It would not be realistic to consider that the simple fact of subjecting aerobic insects to an anoxic environment is sufficient to make them become anaerobic species.

The rice weevil, whose resistance seems to have given rise to the standardized treatment time of three weeks, is an important pest in the agricultural industry. **This insect is not among those frequently encountered in museums.** In museums, the insect population is essentially composed of Anobiidae, Dermestidae, Tineidae, and other families such as the Lyctidae and Lepismatidae (MAEKAWA, S., 1998, PINNINGER, D.). These insects are much less resistant to oxygen deprivation treatment as is indicated by the results of studies involving them.

### *CONCLUSION*

The conclusions we draw after the study carried out with the EPMQB installation are based both on the results of teams abroad working in this field and on our observations at the Musée du quai Branly.

Temperature plays a crucial role for hastening the death of insects found within objects. Thus, at a temperature of 25°C, it is entirely possible to reduce exposure times to 10 or 15 days for the insect species commonly found in museums.

The role played by humidity is less clear-cut in spite of the fact that the principal mechanism leading to insect mortality is desiccation, both for larvae and for adult insects.

In view of the results obtained with the EPMQB installation and also in view of those recorded in the relevant literature for tests on reference samples and on real objects (such as the results of the Getty Conservation Institute), and taking account of the experimental conditions (temperature, relative humidity, oxygen levels) we can move as of now to an exposure time ($T_e$) of 14 days (2 weeks).

The oxygen drop times ($T_i$) being situated between 1 and 2 days for most objects, this corresponds to a treatment time $T_t$ between 15 and 16 days.

$$T_t = T_i + T_e$$

In parallel with the treatment of objects, the rigorous hygiene monitoring programme put in place on the collection treatment site will enable any new infestation to be detected.

For this monitoring programme, the Musée du Quai Branly has called in specialists in tackling infestation. The hygiene monitoring programme involves the setting up of pheromone and baited traps and these traps are renewed every three months.

A programme of harvesting and identification of insects has also been initiated. A list of all the insects found dead or alive, in the treatment sites or on objects is kept by the Cleaning/Dusting Department and transmitted to the Anoxia Department, who are responsible for identification, performed in collaboration with the company Hygiène-Office and the Laboratoire d'Entomologie du Muséum d'Histoire Naturelle[6].

**Bibliography**

BANCK, H., J, ANNIS, P., C., "*Suggested procedures for controlled atmosphere storage of dry grain*". **Commonwealth Scientific and industrial Research Organization ( CSIRO) Division of Entomology Technical Paper, 13.**

BEAR, J., "Hydrodynamic Dispersion", 1963

**KIGAWA, R., *et al*** "*Practical methods of low oxygen atmosphere* and carbon dioxide *treatments for eradication of insect pests in Japan*" in proceedings of 2001: Integrated Pest Management for Collections . A Pest Odyssey. 1-3 October 2001, chapter thirteen.

HOBACK, W.W; STANLEY, D.W. "*Insects in hypoxia*", **J. of Insects Physiology, 2001, 47, 533-544**

PINNINGER, D., "*New Pests for old : The changing status of museum insect pests in the UK*" in **proceedings of 2001: Integrated Pest Management for Collections . A Pest Odyssey. 1-3 October 2001, chapter three.**

RUST, M., VINOD, D., DRUZIK, J., PRESSEUR, F.," *The feasability of using modified atmosphere to control insect pests in museum*", **Restaurator, 1996, 17, 43-60.**

SELWITZ, C. , MAEKAWA, S., "*Inert gases in control of museum insects pests*", **Ed**. **The J. Paul Getty Trust, 1998.**

VALENTIN, N., 1993, "*Comparative analysis of insect control by nitrogen, argon, and carbon dioxide in museum, archive and herbarium collection*", **International Biodeterioration et Biodegradation, 1993, 32, 263-278.**

VALENTIN, N., PREUSSER, F., "*Insect control by inert gases in museums archives and archives*" **Restaurator, 1990, 11, 22-33.**

*Acknowledgements*

We thank Mr. Stéphane Martin, Director of the Musée du quai Branly, for entrusting us with this study.
We also thank the company Hygiène Office for the wooden boxes used in the gas diffusion experiments and the company Mallet for agreeing to modify the installation in line with our needs for this research programme.

---

[6]    Entomology  department of the Natural History Museum, Paris)

We thank Mr. David Pinninger for his assessment of this study, which enabled the Musée du quai Branly to decide if and how it should be implemented.
We thank Mr. Jean-Pierre Mohen, Director of C2RMF[7] for his advice on this project.
We also thank Mr. Jason Hart-Davis for help with the English of this report, and thanks also go to the photographic team of the Musée du quai Branly, and in particular to Ms. Stéphanie Jouane.

---

[7] Centre de Recherche et de Restauration des Musées de France (French Museums Research and Restauration Centre)